\documentclass{article}
\usepackage{graphicx}

\begin{document}

\font\tenbg=cmmib10 at 10pt
\def \rvecmu{{\hbox{\tenbg\char'026}}}
\def \rvecphi{{\hbox{\tenbg\char'036}}}
\def \Omegabold {{\hbox {\tenbg\char'012}}}
\def \rvecOmega {{\hbox {\tenbg\char'012}}}
\font\tenbg=cmmib10 at 10pt
\def \rvecxi{{\hbox{\tenbg\char'030}}}
\def \rvecphi{{\hbox{\tenbg\char'036}}}
\def \rvecdelta {{\hbox {\tenbg\char'016}}}
\def \rvecepsilon {{\hbox {\tenbg\char'017}}}
\def \rvecmu{{\hbox{\tenbg\char'026}}}
\def \Omegabold {{\hbox {\tenbg\char'012}}}
\def \rvecOmega {{\hbox {\tenbg\char'012}}}
\def \cE{{\cal E}}
\def \trho{{\tilde{\rho}}}
\def\lesssim{\mathrel{\hbox{\rlap{\hbox{\lower4pt\hbox{$\sim$}}}\hbox{$<$}}}}
\def\gtrsim{\mathrel{\hbox{\rlap{\hbox{\lower4pt\hbox{$\sim$}}}\hbox{$>$}}}}

\title{Relativistic Poynting-Flux Jets as Transmission Lines}

\author{R.V.E. Lovelace$^{1}$,  S. Dyda$^{2}$, \& P.P. Kronberg$^{3}$\\
$^{1}${\small Department of Astronomy, Cornell University, Ithaca, NY
14853} \\
$^{2}${\small Department of Physics, Cornell University, Ithaca, NY
14853} \\
$^{3}${\small Theoretical Division, Los Alamos Nat. Lab, Los Alamos, NM 87545 and} \\
{\small Department of Physics, University of Toronto, ON M5S 1A7, Canada}\\
{\small {\it email} : lovelace@astro.cornell.edu}
}

\maketitle

\begin{abstract}

    Recent radio emission, polarization,
and Faraday rotation maps of the radio jet of the galaxy 3C 303 have
shown that one knot of this jet has a {\it galactic}-scale electric  current  of $\sim 3\times 10^{18}$ Amp\`ere flowing along the jet axis (Kronberg et al. 2011).
    We develop the theory of relativistic Poynting-flux jets which are
 modeled as a transmission line carrying a DC current $I_0$, 
 having a potential drop $V_0$, and a definite impedance ${\cal Z}_0
 =90(u_z/c)\Omega$, where $u_z$ is the bulk velocity of the jet
 plasma.     The electromagnetic energy flow in the
 jet is ${\cal Z}_0 I_0^2$. 
   The observed current in 3C 303 can be used to 
calculate the electromagnetic
energy flow in this magnetically dominated jet.  
    Time-dependent but not necessarily small
perturbations of a Poynting-flux jet -  possibly triggered
by a gas cloud penetrating the jet -
are described by  ``telegrapher's equations, '' which
predict the propagation speed of disturbances and the 
effective wave impedance ${\cal Z}$.   
   The disturbance of a  Poynting jet by the cloud gives rise
to localized dissipation in the jet which may explain the 
enhanced synchrotron radiation in the knots of the
3C 303 jet, and in the apparently stationary knot HST-1
in the jet from the nucleus of the galaxy M87 (Biretta et al. 1999).

     The formation of a Poynting-flux jet can
be traced back to the dynamics of a large-scale
magnetic field threading the accretion disk around
a black hole.       This magnetic field can arise from the dynamo
processes in the disk triggered by a star-disk 
collision.
   The field may be sufficiently strong  that it
suppresses the magneto-rotational instability with 
the result that the disk is non-turbulent
and without viscosity. 
   The disk will however continue to accrete owing to the angular
momentum outflow in the Poynting-flux  jet.

\end{abstract}

{\bf Introduction:}
Radio emission, polarization,
and Faraday rotation maps of the radio jet of the galaxy 3C303 
(Kronberg et al. 2011) reveal that
 one longitudinal segment of this jet has a {\it galactic}-scale electric 
 current owing along the jet axis. 
     This current can be interpreted as a
relativistic Poynting-flux jet.  
 Here, we develop the theory of relativistic 
Poynting-flux jets by utilizing the
analogy between the jets with   transmission lines.

%%%%%%%%%%%%%%%%%%%%%%%%%%%%%%%%%%%
\begin{figure}[t]
\centering
\includegraphics[scale=0.4]{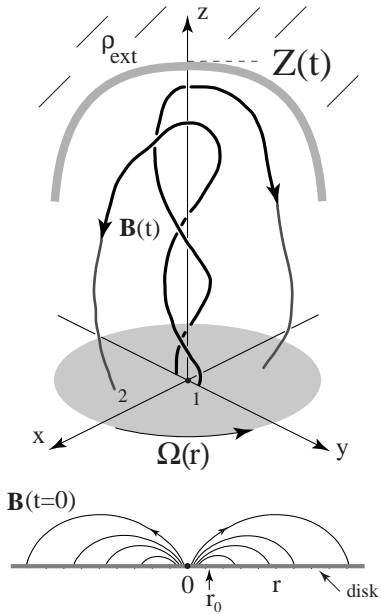}
\caption{Sketch of the magnetic
field configuration of a Poynting jet
adapted from Lovelace and Romanova (2003).
  The bottom part of the figure shows
the initial  dipole-like 
magnetic field threading the disk
which rotates at the angular rate
$\Omega(r)$.  The $O-$point of the
initial field is at $r_0$.
The top part of the
figure shows the jet at some time
later when the head of the jet
is at a distance $Z(t)$. 
   At the head of the jet there is
force balance between electromagnetic
stress of the jet and the ram pressure
of  the ambient medium of density $\rho_{\rm ext}$.}
\end{figure}
%%%%%%%%%%%%%%%%%%%%%%%%%%%%%%%%%%%     

{\bf Theory:}    
    In cylindrical $(r,\phi,z)$
coordinates with
axisymmetry assumed,
    the magnetic field has
the form $ {\bf B}~ = {\bf B}_p +
B_\phi \hat{\rvecphi~},$ with $
{\bf B}_p = B_r{\hat{\bf r}}+
B_z \hat{\bf z}$, and
$B_r =-(1 / r)(\partial \Psi/ \partial z),$ 
and $B_z =(1 / r)(\partial \Psi / \partial r).$
Here,  $\Psi(r,z) \equiv r A_\phi(r,z)$ 
is the flux function.
  A simple form of this function is 
$\Psi(r,0)=(1/2){r^2B_0 /[1+2(r/r_0)^3]}, $
where $B_0$ is the axial magnetic field strength
in the center of the disk, and $r_0$ is
the radius of the $O-$point of the magnetic
field in the plane of the disk as indicated
in Figure 1.    
    This field could arise for $t\geq 0$ from dynamo
processes in the disk triggered by a star-disk 
collision as discussed by Pariev, Colgate, \& Finn (2007).
   The field may be sufficiently strong  that it
suppresses the magneto-rotational instability (MRI) so that the disk is non-turbulent and without viscosity. 
   The disk will however continue to slowly accrete owing to the angular
momentum outflow in the Poynting  jet which gives a
 radial accretion speed  much less than the Keplerian velocity of the
disk.
    This $\Psi$ is taken to apply for
$r \geq 0$ even though it is not valid
near the horizon of the black hole. 
    The contribution
from the latter region is negligble for the
considered conditions where $(r_g/r_0)^2 \ll 1$,
where $r_g\equiv GM/c^2$.
     For a corotating disk around a Kerr black
hole the disk's angular velocity viewed from a large distance is
$\Omega ={[c^3/(GM) ]/[ a_*+(r/r_g)^{3/2}]},$
for $r>r_{\rm ms}$ where $r_{\rm ms}$ is the innermost
stable circular orbit and
$a_*$ is the spin parameter of the black hole
with $0 \leq a_* <1$. 

   At large distances from the disk ($z \gg r_0$) the
the flux function solution of the force-free 
Grad-Shafranov equation is found to be
\begin{equation}
\bar \Psi =  {\bar r^{4/3} \over 
[2 {\cal R}~(\Gamma^2 -1)~]^{2/3}} ~, 
\end{equation}
(Lovelace \& Romanova 2003), where $\Gamma$ is
the Lorentz factor of the jet,
 $\bar r \equiv r/r_0$,
$\bar \Psi \equiv  \Psi/\Psi_0$ with $\Psi_0 \equiv r_0^2 B_0/2$,
and ${\cal R} \equiv r_0/r_g$.
  This dependence holds for
$\bar r_1 \equiv {[2(\Gamma^2-1)]^{1/2}/{\cal R}} < 
\bar r < \bar r_2 =
{[2 {\cal R}(\Gamma^2-1)]^{1/2}/ 3^{3/4}}.$
At the inner radius $\bar r_1$, $\bar \Psi = 1/{\cal R}^2$,
which corresponds to the streamline which passes
through the disk at a distance $r=r_g$.
   For $\bar r < \bar r_1$, we assume
$\bar \Psi \propto \bar r^2$, which
corresponds to $B_z=$ const.
   At the outer radius $\bar r_2$,
$\bar \Psi =(\bar \Psi)_{\rm max} =1/3$ which
corresponds to the streamline which goes
through the disk near the $O-$point at
$r=r_0$.   
    Note that there is an appreciable range of radii
if ${\cal R}^{3/2} \gg 1$.

  For $\bar{r}_1 <r < \bar{r}_2$,  the
field components of the Poynting jet are
\begin{equation}
\bar E_r = -\sqrt{2}~(\Gamma^2-1)^{1/2}~ \bar B_z~,~
\bar B_\phi = -\sqrt{2}~\Gamma~\bar B_z~,
 ~\bar B_z = {2\over 3} {\bar r^{-2/3} 
\over[2{\cal R}(\Gamma^2-1)]^{2/3}}~.
\end{equation}
This electromagnetic field statisfies
the radial force balance equation,
$ {d B_z^2 / dr}+ (1/ r^2){d [ r^2(B_\phi^2 -E_r^2)] /dr} =0.$

   At the jet radius $r_2$,
there is a boundary layer where
the axial magnetic field changes from $B_z(r_2-\varepsilon)$
to zero at $r_2+\varepsilon$,  where $\varepsilon \ll r_2$ is the
half-width of this layer.
     The electric field
changes from $E_r(r_2-\varepsilon)$ to zero at $r_2+\varepsilon$.
    The  toroidal magnetic field
changes from $B_\phi(r_2-\varepsilon)$ to $B_\phi(r_2+\varepsilon)$
where this change is fixed by
the radial force balance.
   Thus for $r>r_2$, we have 
$E_r = 0$, $B_z=0$, and 
$ B_\phi = \sqrt{3} B_z(r_2-\varepsilon)(r_2/r)=
\sqrt{(3/2)}\Gamma^{-1}B_\phi(r_2-\varepsilon)(r_2/r)$.
Equivalently, $B_\phi(r_2+\varepsilon)/B_\phi((r_2-\varepsilon)
=\sqrt{3/2}/\Gamma$.

      The toroidal magnetic field for $r>r_2$ applies out to an
`outer radius' $r_3$ where the magnetic pressure of the jet's
toroidal magnetic field, $B_\phi^2(r_3)/8\pi$,
balances the external ram pressure $P_{\rm ex} = p_{\rm ex}
+\rho_{\rm ex} (dr_3/dt)^2$, where 
$p_{\rm ex}=n_{\rm ex}k_{\rm B}T_{\rm ex}$ is the kinetic
pressure of the external intergalactic plasma and $\rho_{\rm ex}$
is its density.   The outward propagation of the jet will be
accompanied by the non-relativistic expansion the outer radius,
$dr_3/dt >0$.

      We take as the `jet current' the axial current $I_0$
flowing along the jet core $r\leq r_2-\varepsilon$.
  From  Amp\`ere's law, $B_\phi(r_2)  =-2 I_0/(c~\!r_2)$ 
  or in convenient units, $B_\phi[{\rm G}] = -I_0[{\rm A}]/(5 r[{\rm cm}])$.  
  The net current carried by the jet ($r\leq r_2+\varepsilon$)
  is $I_{\rm net} = \sqrt{3/2}\Gamma^{-1} I_0$.
  
    Using equations (2), the energy flux carried by the Poynting jet 
can be expressed as
 \begin{equation}
 \dot{E}_J = {c\over 2}\int_0^{r_2} r dr E_r B_\phi = {\cal Z}_0I_0^2~,
~{\rm where}~~
 {\cal Z}_{0}= {3\over c}\beta
 ~[{\rm cgs}]
 =90 \beta~ \Omega~[{\rm MKS}]~,
\end{equation}
and $\beta=U_z/c = (1-\Gamma^{-2})^{1/2}$.
Here, ${\cal Z}_{0}$ is the DC impedance of the Poynting jet.
 The conversion to MKS units is 
$c^{-1} \rightarrow (4\pi)^{-1}(\mu_0/\epsilon_0)^{1/2} =  30~\! \Omega.$
Earlier, the impedance of a relativistic Poynting jet was estimated
to be $\sim c^{-1}$ (Lovelace 1976).   

For the observed axial current
in the E3 knot in the jet of 3C 303, $3\times 10^{18}$ A 
(Kronberg et al. 2011),  the electromagnetic energy 
flux is $\dot{E}_J \approx 8\times 10^{45}\beta$ erg s$^{-1}$.      
This energy flux is much larger than the photon 
luminosity of the jet of $3.7\times 10^{41}$ erg s$^{-1}$
integrated over $10^8$ to $10^{17}$ Hz (Kronberg et al. 2011) 
assuming $\beta$ is not much smaller than unity.   
    For the E3 knot the jet radius is $r_2 \approx 0.5$ kpc 
so that $B_\phi(r_2) \approx 0.4$ mG.     
   The E3 knot is about $17.7$ kpc from the galaxy nucleus.

{\bf Transmission Line Analogy:}
     Here, we interpret the Poynting jet described 
by equations (6) - (8)  in terms of a transmission
line analogy as suggested by Lovelace \& Ruchti (1983).
  The different physical quantities are measured in
the `laboratory' frame which is rest  frame of the plasma
outside of the jet at $r \geq r_2$.
      The effective potential drop across 
the transmission line  is taken to be 
\begin{equation}
V_0 = - {1\over 2} r_0 \int_0^{\bar{r}_2} d\bar{r} E_r(\bar r) ~~=~~ {r_0 \over 3^{1/4}}
{B_0 \over \sqrt{{\cal R}}}~,
\end{equation}
where the factor of one-half accounts for the fact
that the transmission line does not consist of two
conduction surfaces.

    The axial current flow of the jet is 
\begin{equation}
I_0= - {1\over 2} c r_2 B_\phi(r_2) ={ V_0 \over {\cal Z}_{\rm 0}}~,
\end{equation}
with ${\cal Z}_{\rm 0}$ given by equation (3).
The units of equations (4) and (5) are cgs.
    In MKS units note that a current $I_0 = 3\times 10^{18}$ A
gives a voltage $V_0 = 2.7\times 10^{20}\beta$ V.

 {\bf Electric and Magnetic Field Energies:}  
 The electric field energy per unit length of the
jet in MKS units is
\begin{equation}
W_E ={\epsilon_0 \over 2}2\pi \int_0^{r_2} r dr E_r^2~ = ~{1\over 2}C V_0^2~,
~{\rm where}~~
C={4\pi\epsilon_0 \over 3}~
\end{equation}
is the capacitance per unit length in Farads per meter and
$\epsilon_0 =8.854\times 10^{-12}$ F/m.

    The magnetic energy per unit length of the jet
in MKS units is
\begin{equation}
W_B={\pi\over  \mu_0}\int_0^{r_2} r dr (B_\phi^2 + B_z^2)+
{\pi \over \mu_0} \int_{r_2}^{r_3} r dr B_{\phi+}^2 \left({r_2 \over r}\right)^2~
=~{1\over2}L I_0^2~.
\end{equation}
Here, $B_{\phi+}$ is the toroidal field at $r_2+\varepsilon$.
     Carrying out the integrals we find
\begin{equation}
L ={3 \mu_0 \over 4 \pi}\left[1+{1\over 2 \Gamma^2}
+{1\over 2\Gamma^2}\ln\left({r_3 \over r_2}\right)\right]~,
\end{equation}   
which is the inductance per unit length in Henries per meter
with $\mu_0=4\pi \times 10^{-7}$ H/m.

{\bf Telegrapher's Equations:} Time and space ($z-$)dependent but not necessarily small
perturbations of a Poynting-flux jet are described by
the Telegrapher's equations,
\begin{equation}
{\partial \Delta V \over \partial t}= - {1\over C} {\partial \Delta I \over \partial z}~,
\quad \quad{\partial \Delta I \over \partial t}= - {1\over L} {\partial \Delta V \over \partial z}~,
\end{equation}
where $(\Delta  V, ~\Delta I)$ represent deviations from the equilibrium
values $(V_0,~I_0)$.
The equations can be combined to give the wave equations
\begin{equation}
\left({\partial^2 \over \partial t^2} -u_\varphi^2{\partial^2\over \partial z^2}\right)( \Delta V,~\Delta I) =0~,
\end{equation}
where
\begin{equation}
u_\varphi ={1\over \sqrt{LC}}~=~c\left[1+{1\over 2 \Gamma^2}
+{1\over 2\Gamma^2}\ln\left({r_3 \over r_2}\right)\right]^{-1/2}~,
\end{equation}
is the phase velocity of the perturbation.
   The general solution of equation (21) is
\begin{eqnarray}
\Delta V&=&\Delta V_+(z-u_\varphi t)+\Delta V_-(z+u_\varphi t)~,\nonumber \\
\Delta I&=&\Delta I_+(z-u_\varphi t)+\Delta I_-(z+u_\varphi t)~.
\end{eqnarray}
It is readily shown that
$\Delta V_+ = {\cal Z} \Delta I_+$ and $\Delta V_- =-{\cal Z}\Delta I_-$,
where 
\begin{equation}
{\cal Z} = \sqrt{L\over C} =90 \left[1+{1\over 2 \Gamma^2}
+{1\over 2\Gamma^2}\ln\left({r_3 \over r_2}\right)\right]^{1/2}\Omega ~[\rm MKS].
\end{equation}

{\bf Irregularity in the Transmission Line:}
    The transmission  line may have an  irregularity appear in
it due for example to the intrusion of a plasma cloud at $t > 0$.
 The irregularity  can be modeled as an
 extra impedance ${\cal Z}_\ell$ or `load' across the transmission line at $z=0$.
     This impedance is considered
 to go from ${\cal Z}_\ell(t<0)=\infty$ to a constant value ${\cal Z}_\ell$ for $t>0$.
    In general ${\cal Z}_\ell$ is complex with,
for example, a positive imaginary part if it is dominantly
capacitive.  On either side of the discontinuity the line is
assumed to have a real impedance ${\cal Z}$  given by equation (13).
    On the upstream side of ${\cal Z}_\ell$ ($z<0$), the line voltage
is $V_0 + \Delta V_-$,  where $\Delta V_-$
is the  backward  propagating wave.   
     The current on this part of
the line is $I_0+\Delta I_-$. 
    There is no forward propagating wave for the considered
conditions.
   On the downstream side of ${\cal Z}_\ell$, the line voltage
is $V_0 + \Delta V_t$ and the current is $I_0+\Delta I_t$, where
$(\Delta V_t,~\Delta I_t=\Delta V_t/{\cal Z})$ represents the transmitted
wave.

     The standard conditions on the potential 
and current flow at ${\cal Z}_\ell$ ($z=0$)
give $V_0+\Delta V_- = V_0+\Delta V_t$ and 
$I_0 +\Delta I_- = I_\ell +I_0 +\Delta I_t$, where $I_\ell=
(V_0+\Delta V_-)/{\cal Z}_\ell$
is the current flow through  ${\cal Z}_\ell$.
      In this way we find
 \begin{equation}
 \Delta V_- = {-{\cal Z} \over 2{\cal Z}_\ell +{\cal Z}}V_0~=~\Delta V_t~.
\end{equation}
Note that for ${\cal Z}_\ell \rightarrow \infty$,  both $\Delta V_- $
and  $\Delta V_t $ tend to zero.

     The power loss rate in the load ${\cal Z}_\ell$ is
\begin{equation}
\dot{\cal E}_\ell = {(V_0+\Delta V_t)^2 \over {\cal Z}_\ell}~=~ 
{4 {\cal Z}_\ell \over (2 {\cal Z}_\ell +{\cal Z})^2} V_0^2~.
\end{equation}
We assume that this power goes into accelerating charged particles
which in turn produce the observed synchrotron radiation.
   This power could account for the emission of the E3 knot of
3C 303 (Kronberg et al. 2011) and the emission of the HST-1
knot in the M87 jet (Biretta 1999).

{\bf Acknowledgments:}  RVEL thanks Van Thanh Tran, Roland Triay, Hady Schenten,
and others for organizing this superb meeting.  He also
thanks  Gena Bisnovatyi-Kogan and Marina Romanova for valuable discussions.   RVEL was supported in part by NASA grants  NNX10AF-63G and NNX11AF33G and by NSF grant AST-1008636.  PPK was supported by NSERC
(Canada) grant A5713.

\end{document}